\documentclass[twocolumn,preprintnumbers,amsmath,amssymb]{revtex4}

\usepackage{graphicx}% Include figure files
\usepackage{dcolumn}% Align table columns on decimal point
\usepackage{bm}% bold math
\usepackage{color}% bold math
\begin{document}

\preprint{APS/123-QED}

\title{Magnetoresistance of semi-metals: the case of antimony }

\author{ Beno\^it Fauqu\'e$^{1,2,\footnote{benoit.fauque@espci.fr}}$, Xiaojun Yang$^{2,3}$, Wojciech Tabis$^{4,5}$, Mingsong Shen$^{6}$, Zengwei Zhu$^{6}$, Cyril Proust$^{4}$, Yuki Fuseya$^{2,7}$,}
\author{Kamran Behnia$^{2,\footnote{kamran.behnia@espci.fr}}$}%

\affiliation{
$^1$ JEIP,  USR 3573 CNRS, Coll\`ege de France, PSL Research University, 11, place Marcelin Berthelot, 75231 Paris Cedex 05, France.\\
$^2$ ESPCI ParisTech, PSL Research University CNRS, Sorbonne Universit\'es, UPMC Univ. Paris 6, LPEM, 10 rue Vauquelin, F-75231 Paris Cedex 5, France \\
$^3$ School of Physics and Optoelectronics, Xiangtan University, Xiangtan 411105, China.\\
$^4$ Laboratoire National des Champs  Magn\'etiques Intenses (LNCMI-EMFL), CNRS ,UGA, UPS, INSA, Grenoble/Toulouse, France\\
$^5$ AGH University of Science and Technology, Faculty of Physics and Applied Computer Science, 30-059 Krakow, Poland\\
$^6$Wuhan National High Magnetic Field Center and School of Physics, Huazhong University of Science and Technology, Wuhan 430074, China \\
$^7$Department of Engineering Science, University of Electro-Communications, Chofu, Tokyo 182-8585, Japan\\
}

\date{\today}% It is always \today, today,

\begin{abstract}
Large unsaturated magnetoresistance has been recently reported in numerous semi-metals. Many of them have a topologically non-trivial band dispersion, such as Weyl nodes or lines. Here, we show that elemental antimony displays the largest high-field magnetoresistance among all known semi-metals. We present a detailed study of the angle-dependent  magnetoresistance and use a semi-classical framework invoking an anisotropic mobility tensor to fit the data. A slight deviation from perfect compensation and a modest variation with magnetic field of the components of the mobility tensor are required to attain perfect fits at arbitrary strength and orientation of magnetic field in the entire temperature window of study. Our results demonstrate that large orbital magnetoresistance is an unavoidable consequence of low carrier concentration and the sub-quadratic magnetoresistance seen in many semi-metals can be attributed to field-dependent mobility, expected whenever the disorder length-scale exceeds the Fermi wavelength.
\end{abstract}

\pacs{Valid PACS appear here}% PACS, the Physics and Astronomy
\maketitle

Detailed studies of magnetoresistance have been carried out in a large variety of systems ranging from ferromagnetic-metallic multilayer devices \cite{Levy94}, manganite perovskites \cite{Rao98}, and doped semi-conductors \cite{Solin00}. The large increase in the resistivity of semi-metals, which appears when the magnetic field is applied perpendicular to the orientation of the electric current has been known for a long time \cite{Kapitza28}. Recently, such a large orbital magnetoresistance has been observed in many dilute metals. The list includes three-dimensional Dirac systems (Cd$_3$As$_2$ \cite{Liang2014}), Weyl semi-metals (WTe$_2$ \cite{Ali14}, NbP \cite{Shekhar15} or WP$_2$ \cite{Kumar17,Schonemann2017}, LaBi \cite{Shanshan16}), but also 'trivial' semi-metals (LaSb \cite{Tafti2016}, gray arsenic\cite{Zao2017}). Reporting on the large magnetoresistance in WTe$_{2}$, Ali \emph{et al.} \cite{Ali14} contrasted its unsaturated B$^{2}$ behavior to the quasi-linear and eventually saturating high-field magnetoresistance of well-known semi-metals such as bismuth and graphite. Soon afterwards, WTe$_{2}$ was identified as a type-II Weyl semimetal \cite{Soluyanov15}. During the last three years, the link between amplitude and the field-dependence of magnetoresistance and its possible link to non-trivial electronic topology became a subject of debate and motivated numerous studies.

Here, we present the case of elemental antimony, another well-known semi-metal.  Despite the remarkably high mobility of its bulk carriers \cite{Aleksandrov72} and its topologically non-trivial surface states\cite{Seo10}, its large bulk magnetoresistance \cite{Bresler72,Brandt68} has never been studied in detail. Its carrier density of n$\simeq$p$\simeq$5.5 $\times$ 10$^{19}$ cm$^{-3}$ \cite{Windmiller66} is comparable to WTe$_2$ (n$\simeq$p$\simeq$6.6$\times$ 10$^{19}$cm$^{-3}$ \cite{Zhu15}) and two orders of magnitude larger than bismuth (n$\simeq$p$\simeq$3$\times$ 10$^{17}$cm$^{-3}$ \cite{Liu95}). We show that in a reasonably clean Sb single crystal, a magnetic field of 56 T enhances resistivity by a factor of  $\frac{\Delta \rho}{\rho_0}= 3\times$ 10$^6$. This is one order of magnitude higher than what was reported in WTe$_2$ \cite{Ali14}, and comparable with what was observed in WP$_2$ \cite{Kumar17}. The field dependence of magnetoresistance is close to (but slightly different from) quadratic. We present an angle-dependent study, with magnetic field kept perpendicular to the applied current and rotating in three perpendicular crystalline planes (as in the case of bismuth \cite{Collaudin15}), and employ a semiclassical picture, based on distinct mobility tensors for electrons and holes, to explain the data. We show that by assuming slightly imperfect compensation, one can explain the finite Hall response. Taking into account a smooth field-induced reduction in mobility allows  perfect fits to the non-quadratic magnetoresistance  and the non-linear Hall resistivity at the same time. Thus, antimony becomes the first semi-metal whose extremely large magnetoresistance is totally explained by an extended semiclassical treatment for any arbitrary amplitude and orientation of magnetic field.

The two additional assumptions required to attain perfect fits bring new insight to the apparent diversity of semi-metallic magnetoresistance. Imperfect compensation caused by 1 part per million (ppm) of uncontrolled doping is barely noticeable when there is one hole and one electron per 10$^4$ atoms, that is when carrier density is in the range of 10$^{19}$cm$^{-3}$. But the same amount of uncontrolled doping has a much stronger signature when the carrier density is two orders of magnitude lower. This explains why the high-field magnetoresistance is close to quadratic in Sb and WTe$_{2}$, but almost linear in Bi. The field-induced reduction in mobility can be caused by a disorder invisible to long-wavelength electrons at zero-field and becoming relevant in presence of magnetic field.  Since the electron wave-function is smoothly squeezed by magnetic field,  Raleigh scattering by extended defects becomes more efficient as the Landau levels are depopulated \cite{Song15}. This provides a simple, but non-universal foundation for the field-induced decrease in mobility leading to the ubiquitous non-quadratic magnetoresistance.

\begin{figure}[]
\begin{center}
\includegraphics[angle=0,width=8.5cm]{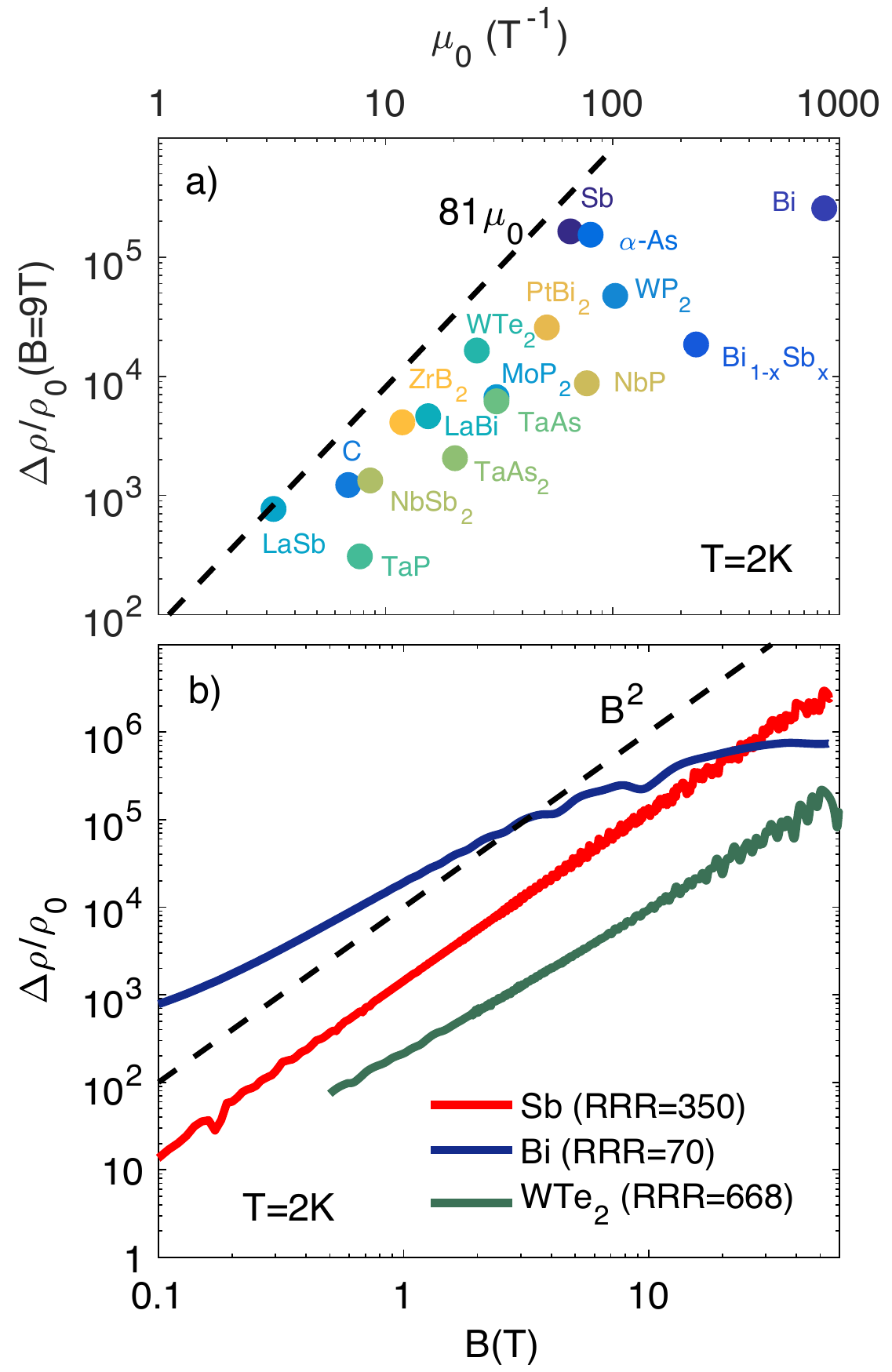}
\caption{a) Magnetoresistance of various semi-metals at B=9T and T=2K as a function of the mobility $\mu_0$=$\frac{1}{\rho_0(n+p)e}$ where $e$ is the charge of the electron,  $n$ and $p$  are the electron and hole density and $\rho_0$ the zero field resistivity at 2K. $\mu_0$ is expressed in Tesla$^{-1}$=10$^4$cm$^2$.V$^{-1}$.s$^{-1}$ b) Magnetoresistance of elemental semi-metals antimony (j//trigonal and B//Bisectrix), bismuth (j//bisectrix and B//trigonal) and WTe$_2$ (j//a-axis and B//c-axis) at T=2K. RRR (Residual Resistivity Ratio) is equal to $\frac{\rho(T=300K)}{\rho(2K)}$.}.
\label{Fig1}
\end{center}
\end{figure}

Fig. 1a) presents the reported magnetoresistance of numerous semimetals at B=9T and T=2K (See \cite{Supplement} for details). The amplitude of field-induced enhancement in resistivity is plotted as a function of zero-field mobility, extracted from resistivity and carrier density: $\mu_0$= $\frac{1}{\rho_0(n+p)e}$. Here, $e$ is the charge of the electron,  $n$ and $p$  are the electron and hole densities and $\rho_0$ the zero-field resistivity at T=2K. One can see that across more than three orders of magnitude, the MR of semi-metals (topological or not) roughly scales with their zero-field mobility. The higher the mobility, the larger the magnetoresistance. However, one can, also see that systematically $\frac{\Delta \rho}{\rho_0}$ is lower than 81$<\mu_{0}^2>$, which is what is expected for 9 T if the mobility was the same unique number relevant to the two (zero-field conductivity, $\sigma$=$e(n\mu_e+p\mu_h)$, and  finite-field magnetoresistance).

Even the most ideally simple semimetal requires more complexity. Such a system would have a single electron-like and a single hole-like Fermi surface. The two pockets are isotropic with scalar mobilities of $\mu_{e}$  and  $\mu_{h}$. The semi-classic magnetoresistance of such a system would be $\frac{\Delta \rho}{\rho_0}=<\mu^2_m> B^2$, where $\mu_m=\sqrt{\mu_e\mu_h}$. Note that the zero-field average mobility is $\mu_0=\frac{n\mu_e+p\mu_h}{(n+p)}$. The two average mobilities are identical only when n=p and $\mu_e=\mu_h$. However, $\mu_0$ and $\mu_m$ remain of the same order of magnitude.

\begin{figure*}[]
\begin{center}
\includegraphics[angle=0,width=16cm]{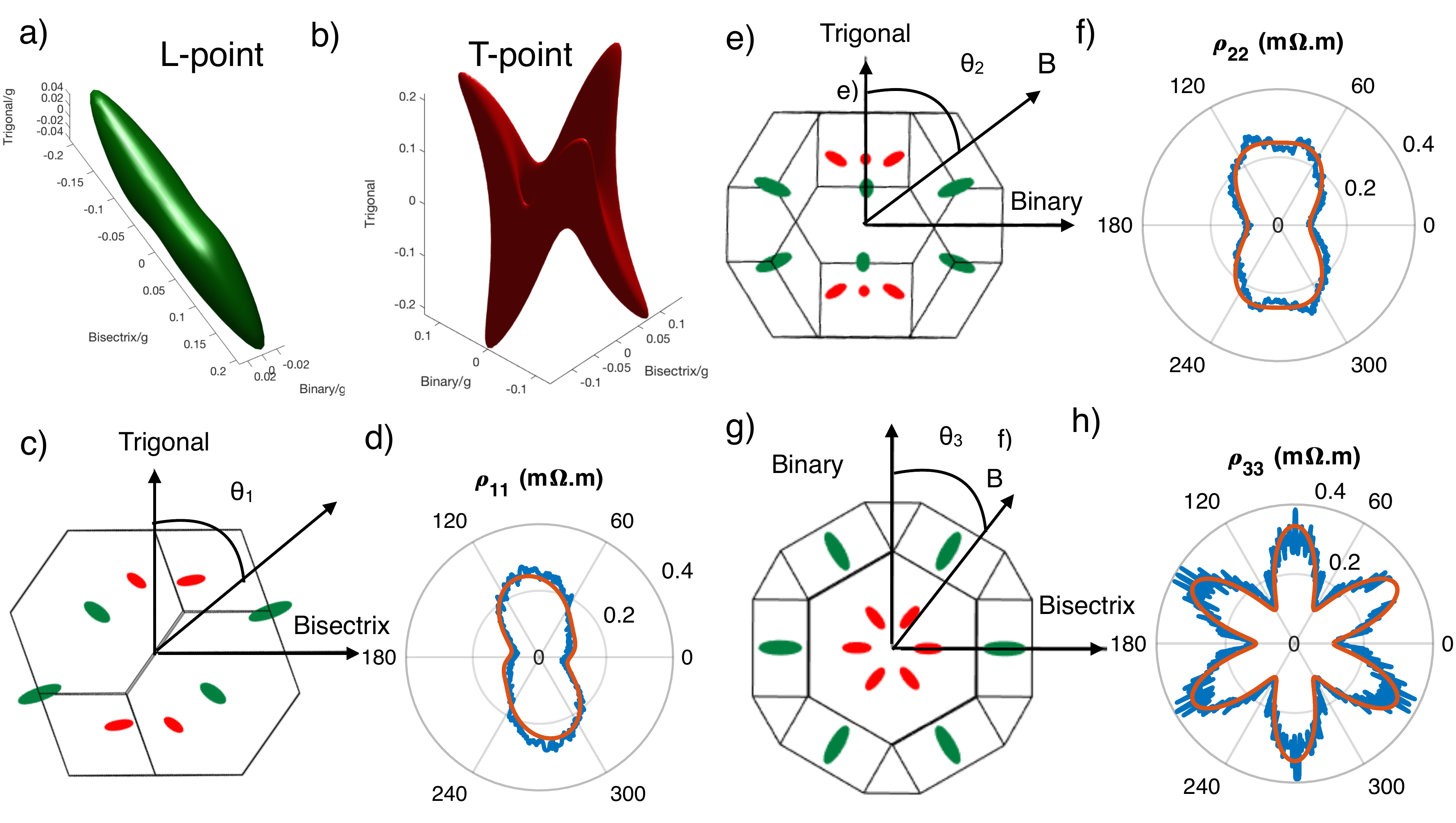}
\caption{Angular dependence of the magnetoresistance of antimony.  a)-b) Electron and hole Fermi surface of antimony according to the Liu and Allen tight binding model. The unit of reciprocal length is $g=1.461 $\AA$^{-1}$. c) e) g) Sketch of the Brillouin zone and the Fermi surface of antimony projected in the three planes of rotation P$_1$ (Trigonal, Bisectrix), P$_2$ (Trigonal, Binary) and P$_3$ (Binary, Bisectrix). d) f) h)  $\rho_{ii}$ for $i$=1,2 and 3 in the three planes of rotation P$_1$, P$_2$ and P$_3$ in blue at T=2K and B=12T (the electric current is applied along the crystal axis perpendicular to the rotating plane) measured on the same sample. The red line is a fit using a theoretical model described in the text.}
\label{Fig2}
\end{center}
\end{figure*}

As seen  in Fig. 1a, despite the discovery of numerous new compounds, the three elemental semimetals (Bi, Sb and As) are still those showing the largest magnetoresistance at 9 T.  Fig. \ref{Fig1}b) compares the field dependence of magnetoresistance in bismuth, WTe$_2$ and antimony at T=2K and B$\leq$56 T. One can see that it is close to B$^2$ in Sb and WTe$_2$,  but presents a lower exponent (B$^{1.5}$) and a tendency to saturation in bismuth. As a consequence, antimony surpasses bismuth above $\approx$ 25T. Note that the amplitude of magnetoresistance in a given metal is not fixed and depends on the cleanness as we will discuss below.

A more elaborate picture requires one to consider the tensorial nature of mobility. We attempted to achieve this by studying the angle dependence of $\rho_{ii}$  in Sb when the magnetic field rotates in the plane perpendicular to the i-axis. Here, ($i$=1,2,3) refer to the binary, bisectrix and trigonal crystal axes. The three planes of rotations are P$_1$=(trigonal, bisectrix), P$_2$=(trigonal, binary) and P$_3$=(binary, bisectrix), as in the case of bismuth \cite{Collaudin15}.

The Fermi surface of Sb has been intensively studied by quantum oscillations \cite{Brandt68,Windmiller66} and cyclotron resonance \cite{Datars64}. These studies have found that the density of electron and hole pockets is equal to $n$=$p$=5.5$\times10^{19}$cm$^{-3}$, within a precision of 1$\%$\cite{Windmiller66}. In these studies, the Fermi surface was taken to be three equivalent electron pockets and six equivalent hole pockets. The electron pockets are quasi ellipsoids centered at the L-points of the Brillouin zone and reminiscent of the case of  bismuth. The hole pockets consist of two groups of three pockets slightly off the  T-points of the Brillouin zone and therefore non-isomorphic to the case of bismuth. According to the tight-binding model used by Liu and Allen \cite{Liu95}, as discussed in the S. I. section C \cite{Supplement}, the six hole pockets are interconnected around the T-point, as shown in Fig. 2b, a feature not explicitly mentioned previously \cite{Liu95}.

The angle-dependence of the magnetoresistance at T=2K and B=12T measured on the same sample for the three perpendicular planes of field rotation are presented in Fig.2d, f, h. In each case, the profile of $\rho_{ii}(\theta)$  reflects the symmetry of the projected profile of the Fermi surface components in that plane (Fig.2c, e, g). When the current is along the binary and the magnetic field rotates in the P$_1$ plane, there is only the inversion symmetry so that : $\rho_{11}(\theta)$=$\rho_{11}(\pi+\theta)$. When the current is along the bisectrix and a magnetic field rotates in P$_2$ plane, as a consequence of additional mirror symmetry, $\rho_{22}(\theta)=\rho_{22}(-\theta)$. Finally, when the current is along the trigonal and the field rotates in the P$_3$  plane, sixfold oscillations results as a consequence of the C$_3$ symmetry combined with the inversion symmetry. In contrast to bismuth \cite{Zhu12,Collaudin15} we did not observe any field-induced loss of threefold symmetry (see section E of \cite{Supplement} for more details).

In order to extract the components of the mobility tensor for electrons and holes, we used the formalism developed first by the Mackey and Sybert \cite{Mackey69} and  then by Aubrey \cite{Aubrey}. In this approach, the total conductivity is the sum of the contributions by each valley expressed as:

\begin{equation}\label{1}
\hat{\sigma}_{e,h} = ne \ ( \hat{\mu}_{e,h}^{-1} \mp \hat{B} )^{-1}
\end{equation}

where $\hat{B}$ is defined as:
\begin{equation}\label{2}
\hat{B}= \begin{pmatrix} 0 & -B_3 & B_2 \\ B_3 & 0 & -B_1 \\ -B_2 & B_1 & 0 \end{pmatrix}
\end{equation}
Here, $\hat{\mu}_{e,h}$ is the  mobility tensor of electrons and holes.  B$_{1}$, B$_{2}$ and B$_{3}$ are the projections of the magnetic field vector along the three principal axes. By employing this formalism, we could extract the components of the mobility tensor along the principal axes of the representative ellipsoids for electrons, $\mu_{i}$, and holes, $\nu_{i}$(i=1, 2, 3). Further details can be found in \cite{Supplement} section D.

\begin{figure}[]
\begin{center}
\includegraphics[angle=0,width=8.9cm]{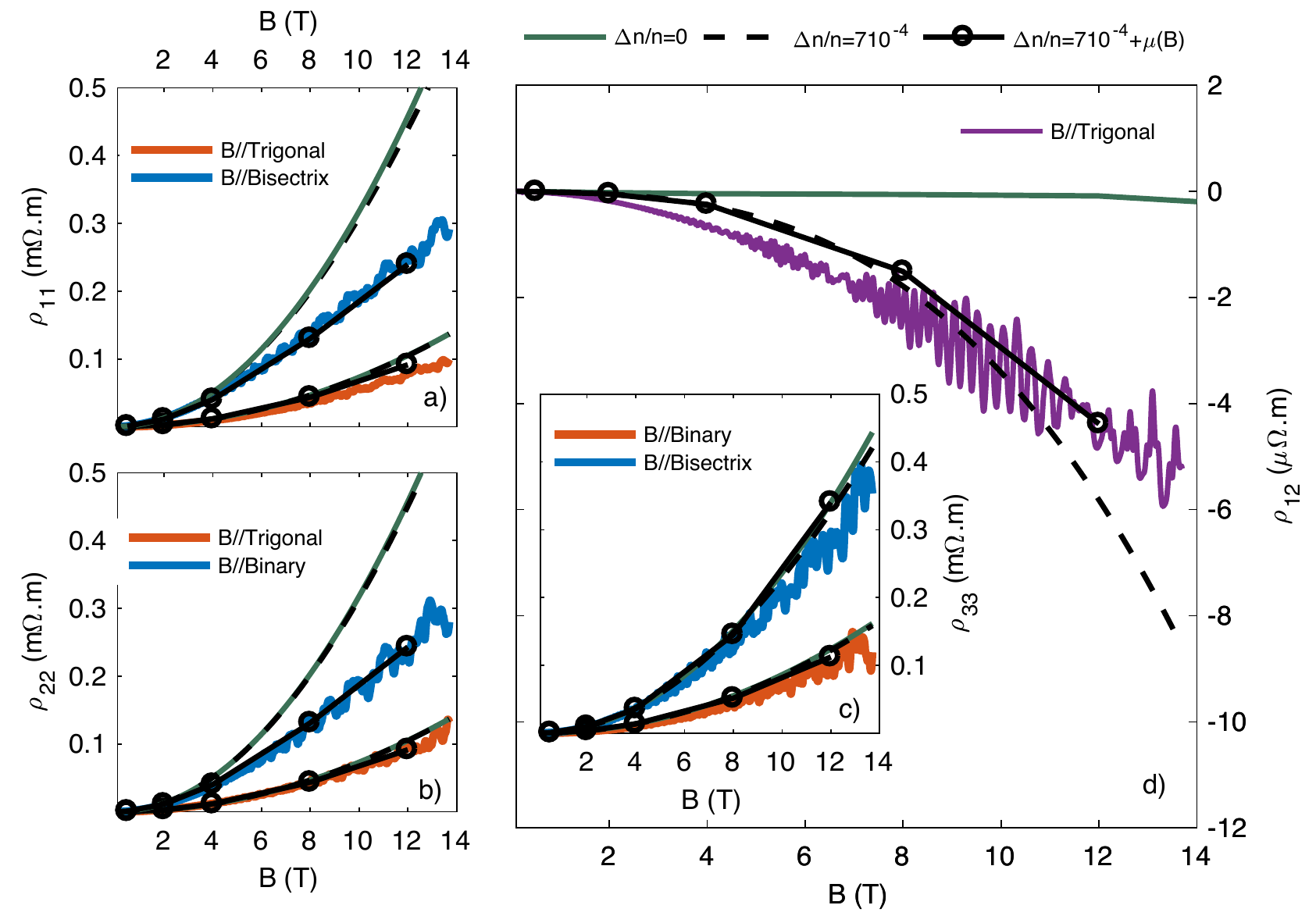}
\caption{Transverse magnetoresistance of Sb at T=2K along the axes of high symmetry : a) $\rho_{11}$(B) for B// trigonal (in blue) and B//bisectrix (in red), b) $\rho_{22}$(B) for B// trigonal (in blue) and B//binary (in red) c)  $\rho_{33}$(B) for B//binary (in blue) and B//bisectrix (in red), d) $\rho_{12}$ for B//trigonal (in purple). The green lines correspond to the Mackey and Sybert's model assuming no change in mobility tensors with the magnetic field and a perfect compensation ($\frac{\Delta n}{n}$=0). The dashed black lines correspond to the same model with a deviation from perfect compensation equal to $\frac{\Delta n}{n}$=7$\times$10$^{-4}$ which allow to capture the Hall response. The black lines correspond to the same model with a field dependence mobility tensor (reported on Fig. 4b) and a deviation from perfect compensation $\frac{\Delta n}{n}$=7$\times$10$^{-4}$.}
\label{FigSweepB}
\end{center}
\end{figure}

We needed to add two ingredients to this formalism in  order to substantially improve the fits. These were a slight departure from perfect compensation and taking into consideration the evolution of $\mu_{i}$ and $\nu_{i}$ with magnetic field. In this formalism, the Hall response is expected to be linear in the high-field  regime ($\mu B\gg 1$) if the compensation is perfect.  As illustrated in Fig.\ref{FigSweepB} d), experimentally, this is not the case. The non-linear Hall resistivity can be captured by assuming a slight imbalance between hole and electron density, $\frac{\Delta n}{n}=\frac{p-n}{n}$.  We find that the best agreement with data yields $\frac{\Delta n}{n}=7\times10^{-4}$. This roughly corresponds to one defects per million unit cell, consistent with the 5N quality Sb powder used to grow the sample and  with early galvanometric measurements \cite{Bresler72}. The consistency of our analysis is checked by comparing the field dependence of $\rho_{11}$, $\rho_{22}$ and $\rho_{33}$ (See Fig.\ref{FigSweepB} a-d). One can see that imperfect compensation combined with a slight field-dependence of the components of the mobility tensor provides a satisfactory description of the overall field dependence of $\rho_{11}$, $\rho_{22}$, $\rho_{33}$ and $\rho_{12}$.

\begin{figure}[]
\begin{center}
\includegraphics[angle=0,width=8.5cm]{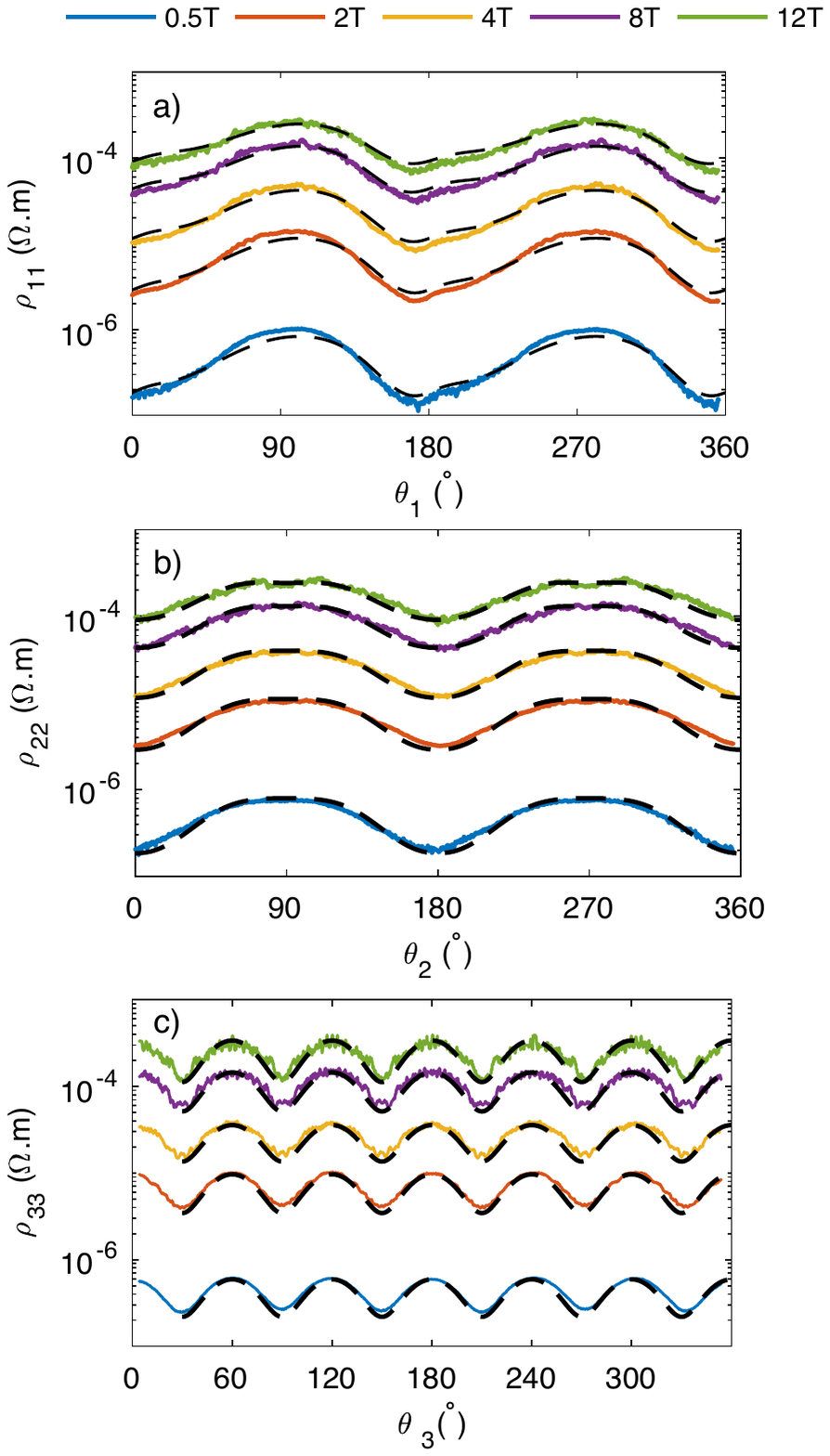}
\caption{Angular dependence of the magnetoresistance of Sb at T=2K for B=0.5T to 12T in the three plans of rotations a) P$_1$, b) P$_2$ and c) P$_3$. The dot lines are fits using the semi-classical model describes in the text.}
\label{FigRot2K}
\end{center}
\end{figure}

\begin{figure}[]
\begin{center}
\includegraphics[angle=0,width=8.5cm]{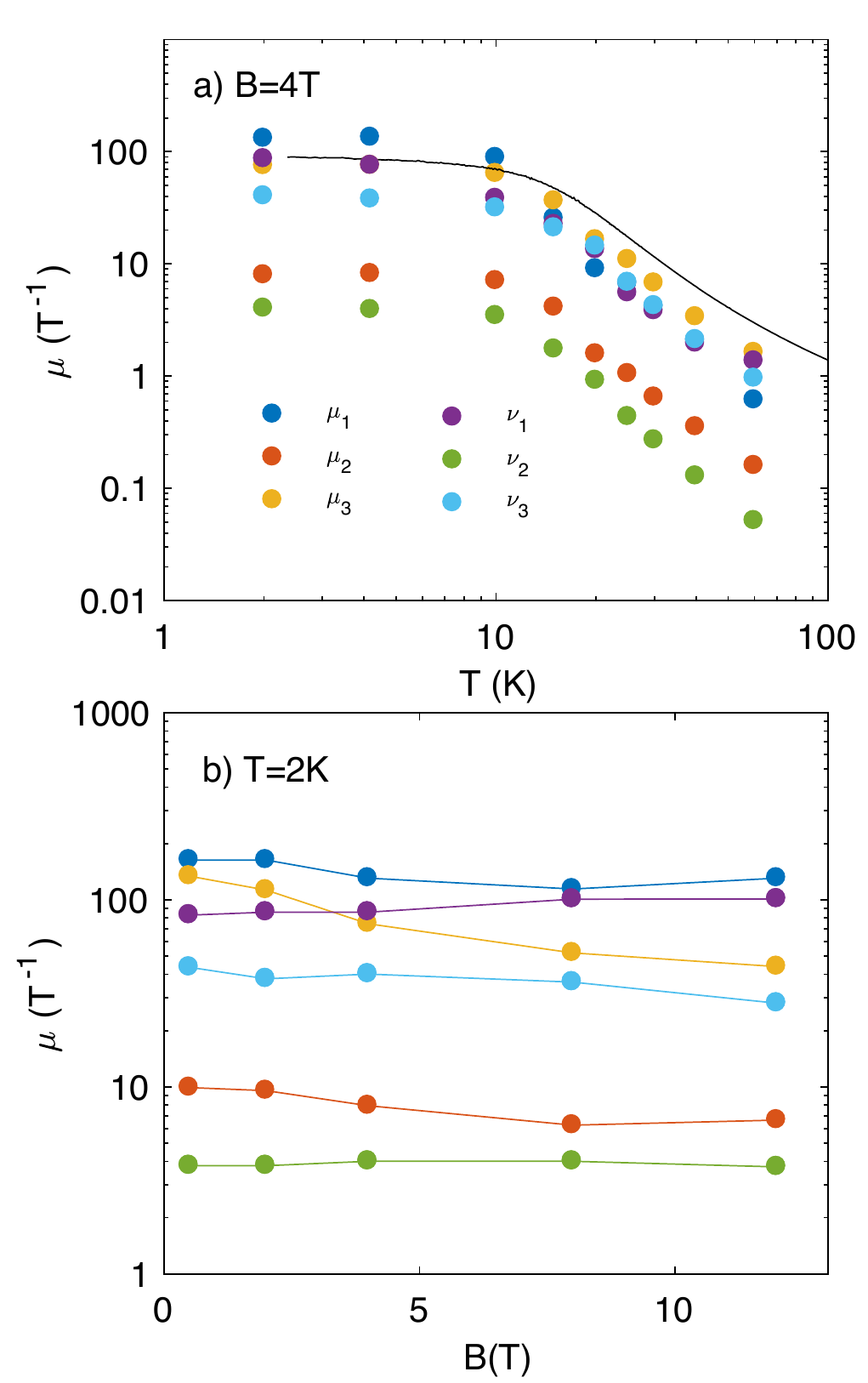}
\caption{Temperature and field dependence of mobility tensors of Sb : the hole and electron tensor component are respectively labelled ($\nu_1$,$\nu_2$,$\nu_3$), ($\mu_1$,$\mu_2$,$\mu_3$) a) $\nu_i$ and $\mu_i$ for $i$=1,2,3 at B=4T as function of temperature. The black line is the mobility deduced from the temperature dependence of $\rho_{33}$ in the Drude picture b) Field dependence of $\mu_i$ and $\nu_i$ for $i$=1,2,3 at T=2K deduced from the fit reported on Fig. \ref{FigRot2K}}
\label{FigTemp}
\end{center}
\end{figure}

Our best simultaneous fits of $\rho_{11}$, $\rho_{22}$ and $\rho_{33}$ at T=2K and B=12T are represented in red in Fig.\ref{Fig2}. We performed similar fits at different temperature and magnetic fields  (see Fig.\ref{FigRot2K}) to obtain the magnitude of the components of the mobility tensors at any magnetic field and temperature in Fig.\ref{FigTemp}.  One can see that the temperature dependence closely tracks the temperature dependence of the zero-field mobility along the trigonal axis: $\mu_{0}[\rho_{33}]$=$\frac{1}{ne\rho_{33}(B=0)}$.  At low temperature, the components of the mobility tensor saturate. Above  10 K,  they show a  T$^{-2}$,  like the one seen in bismuth\cite{Collaudin15} and attributed to inter-valley carrier-carrier scattering\cite{Hartman69}.

The mobility tensors and the mass tensors are linked by $\hat{\mu}_{e,h} =e \hat{\alpha}_{e,h}\cdot\hat{\tau}_{e,h}$. Here, $\hat{\alpha}$ and $\hat{\tau}$ share the same principal axes and $\hat{\alpha}$ is the inverse of the mass tensor. Its components are the cyclotron masses \cite{Datars64}. The anisotropy of the scattering rate is set by (and is attenuated compared to) the anisotropy of the effective mass(see section D of \cite{Supplement}).

The field dependence of the components of the mobility tensor is also instructive. When the field is swept from 0.5T to 12T, $\mu_1$, $\nu_1$, $\nu_2$ are almost constant, but $\mu_2$ and $\nu_3$ decrease by a factor of 1.5 and $\mu_3$ by a factor of 3.  The fact that magnetic field does not affect in a similar manner among different components of the mobility tensor is a clue to the origin of the scattering process which becomes more effective in presence of magnetic field. The drastic decrease in $\mu_3$ indicates  that disorder in the orientation perpendicular to the cleaving planes becomes more significant with increasing field. Another piece of evidence is provided by contrasting samples with different levels of disorder. As presented in \cite{Supplement}, the sub-quadratic aspect of magnetoresistance becomes more pronounced in dirtier samples where both the RRR (Residual Resistivity Ratio) and the Dingle mobility, $\mu_D$ are lower. A ratio of $\frac{\mu_0}{\mu_D}$ as large as 300 implies a significant role for small angle scattering. We thus identify long-range disorder as the ultimate source of sub-quadratic magnetoresistance.

In conclusion, we found a very large magnetoresistance in elemental antimony towering over most reported cases. We showed that the magnitude of the magnetoresistance at any temperature and magnitude or orientation of the magnetic field can be explained using a semi-classical framework with two additional ingredients, a weak deviation from perfect compensation and a field-dependent mobility tensor. The same framework can be used to address  all other semi-metals. We conclude that because of slight imperfect compensation sub-quadratic magnetoresistance is unavoidable. In  antimony or in WTe$_2$, the carrier concentration is large enough to cover the role played by imperfect compensation, yet small enough to allow carriers to have a large mobility. This is the main reason behind its towering magnetoresistance. The phenomenological achievement reported here is only one step towards a microscopic theory.  The challenge is to describe the anisotropic effect of magnetic field on carriers with a known dispersion in presence of a precise scattering potential. This would pin down the intrinsic limits of mobility in a given material.

\section*{Acknowledgments}
This work is supported by the Agence Nationale de Recherche as a part of the QUANTUMLIMIT project, by the Fonds-ESPCI and  by a grant attributed by the Ile de France regional council. We acknowledge support from the LNCMI, a member of the European Magnetic Field Laboratory(EMFL). BF acknowledges support from Jeunes Equipes de l'Institut de Physique du Coll\`ege de France (JEIP). YF is supported by JSPS KAKENHI grants 16K05437 and 15KK0155. We thank D. Carpentier for stimulating discussions.

\end{document}